 \documentclass[preprint,showpacs,nofootinbib,aps]{revtex4}
 \usepackage{graphicx}
 \usepackage{bm}
 
\begin{document}
 \title{Study of $B_{s}$ ${\to}$ ${\eta}_{c}(J/{\psi})D$ decay
        with perturbative QCD approach}
 \author{Junfeng Sun}
 \affiliation{College of Physics and Electronical Engineering,
              Henan Normal University, Xinxiang 453007, China}
 \author{Zhengjun Xiong}
 \affiliation{College of Physics and Electronical Engineering,
              Henan Normal University, Xinxiang 453007, China}
 \author{Yueling Yang}
 \affiliation{College of Physics and Electronical Engineering,
              Henan Normal University, Xinxiang 453007, China}
 \author{Gongru Lu}
 \affiliation{College of Physics and Electronical Engineering,
              Henan Normal University, Xinxiang 453007, China}

 \begin{abstract}
 The $W$-exchange process $B_{s}$ ${\to}$ ${\eta}_{c}(J/{\psi})D$
 is studied with the perturbative QCD approach. Three kinds of
 wave functions for $B_{s}$ meson and two forms of wave functions
 for charmonium are considered. It is estimated that branching
 ratios for $B_{s}$ ${\to}$ ${\eta}_{c}\overline{D}$,
 ${\eta}_{c}D$, $J/{\psi}\overline{D}$, $J/{\psi}D$ decays
 are the order of $10^{-7}$, $10^{-8}$, $10^{-8}$, $10^{-9}$,
 respectively, where the largest uncertainty is from wave functions.
 There is a possibility for measuring these decay in the near future.
 \end{abstract}
 \pacs{13.25.Hw 14.40.Nd}
 \maketitle

 The first evidence for $B_{s}$ production in $e^{+}e^{-}$
 annihilation at the ${\Upsilon}(5S)$ resonance was found
 by the CLEO collaboration \cite{prl.96.022002}.
 Belle has accumulated 121 fb${}^{-1}$ data at the
 ${\Upsilon}(5S)$ resonance, including $7.1{\times}10^{6}$
 $B_{s}\overline{B}_{s}$ pairs \cite{arxiv:1212.5342}.
 It is estimated that some $5.9{\times}10^{8}$ $B_{s}$ mesons
 in the dataset of 5 ab${}^{-1}$ at the ${\Upsilon}(5S)$
 resonance in a New Snowmass Year (about $10^{7}$ seconds
 of actual annual running time \cite{arXiv:0709.0451})
 will be collected at the high luminosity $e^{+}e^{-}$
 asymmetric SuperKEKB \cite{arxiv:1002.5012}.
 More and more $B_{s}$ decays will have subjected the
 Standard Model and new physics to a series of increasingly
 stringent tests, through observables such as branching ratios,
 CP-violationg asymmetries and kinematic distributions.

 In the naive spectator model, the general properties of
 the $B_{s}$ meson parallel those of the $B_{u,d}$ mesons.
 The close correspondences between $B_{s}$ and $B_{u,d}$
 mesons allow for sensitive tests of hadronic models
 \cite{arxiv:1212.5342}.
 Hadronic $B$ decays are complicated on account of strong
 interaction effects, meanwhile, they will have provided a
 great opportunity to study perturbative and nonperturbative
 QCD.
 For nonleptonic two-body $B$ decays, the low-energy
 effective Hamiltonian ansatz and the factorization
 hypothesis are commonly used.
 In recent years, several attractive methods have been proposed
 and widely used to evaluate the hadronic matrix elements
 (where the local operators in the effective Hamiltonian are
 sandwiched between initial and final hadron states considered)
 based on an expansion in the QCD coupling constant
 ${\alpha}_{s}/{\pi}$ and in the power ${\Lambda}_{\rm QCD}/m_{Q}$
 (where ${\Lambda}_{\rm QCD}$ and $m_{Q}$ are the QCD
 characteristic scale and the mass of heavy quark $Q$,
 respectively), such as the QCD factorization \cite{qcdf},
 perturbative QCD method (pQCD) \cite{pqcd},
 soft and collinear effective theory (SCET) \cite{scet}, etc.

 Using the operator product expansion and renormalization group
 equation, the low energy effective Hamiltonian for the
 $B_{s}$ ${\to}$ ${\eta}_{c}(J/{\psi})D$ decay can be
 written as \cite{rmp68p1125}:
  \begin{eqnarray}
 {\cal H}_{eff}
 &=& \frac{G_{F}}{\sqrt{2}} V_{cb}^{\ast}V_{us} \Big\{
    C_{1}(\bar{b}_{\alpha}c_{\alpha})_{V-A}(\bar{u}_{\beta}s_{\beta})_{V-A}
  + C_{2}(\bar{b}_{\alpha}c_{\beta})_{V-A}(\bar{u}_{\beta}s_{\alpha})_{V-A}
     \Big\} \nonumber\\&+&
     \frac{G_{F}}{\sqrt{2}} V_{ub}^{\ast}V_{cs} \Big\{
    C_{1}(\bar{b}_{\alpha}u_{\alpha})_{V-A}(\bar{c}_{\beta}s_{\beta})_{V-A}
  + C_{2}(\bar{b}_{\alpha}u_{\beta})_{V-A}(\bar{c}_{\beta}s_{\alpha})_{V-A}
     \Big\} + \hbox{H.c.} ,
  \label{eq:hamiltonian}
  \end{eqnarray}
 where $G_{F}$ is the Fermi coupling constant.
 The Cabibbo-Kobayashi-Maskawa (CKM) matrix factors
 $V_{cb}^{\ast}V_{us}$ (or $V_{ub}^{\ast}V_{cs}$) and
 the Wilson coefficients $C_{1,2}$ describe the strengths
 of the local four-quark operators in the effective
 Hamiltonian.
 ${\alpha}$ and ${\beta}$ are $SU(3)$ color indices.
 $(\bar{q}q^{\prime})_{V-A}$ $=$
 $\bar{q}{\gamma}_{\mu}(1-{\gamma}_{5})q^{\prime}$.
 The Wilson coefficients, which incorporate the physics
 contributions from high scales, have been calculated
 to the next-to-leading order in the perturbation
 theory and evolved to a characteristic scale with
 the renormalization group equation \cite{rmp68p1125}.
 The essential problem obstructing the calculation of
 nonleptonic decay amplitudes is how to evaluate the
 hadronic matrix elements of the local operators
 properly and accurately.

 Based on the principle of color transparency \cite{color}
 and factorization scheme, the phenomenological treatment
 of the hadronic matrix elements for the $W$-exchange
 processes $B_{s}$ ${\to}$ ${\eta}_{c}(J/{\psi})D$ is
 the same as that for pure annihilation topologies.
 Although the annihilation amplitude is formally power
 suppressed by ${\Lambda}_{\rm QCD}/m_{b}$ with the QCDF power
 counting conventions \cite{qcdf}, its contribution is
 indispensable for realistic $B$-meson decays \cite{pqcd}.
 The comprehensive analysis of $B_{u,d}$ ${\to}$ $PP$,
 $PV$ decays without taking into account the annihilation
 contributions is generally of poor quality \cite{du}
 ($P$ and $V$ denote the pseudoscalar and vector mesons,
 respectively).
 Study of $B_{s}$ ${\to}$ ${\eta}_{c}(J/{\psi})D$
 decay will help to improve our understanding of the
 annihilation effects.

 Analogous to the analysis for hard exclusive scattering
 amplitudes \cite{prd.22.2157}, the hadronic matrix
 element is commonly expressed as a convolution of
 scattering kernels with the universal wave functions
 of the participating hadrons \cite{qcdf,pqcd},
 where nonperturbative dynamics either cancel or is
 absorbed into hadron wave functions (WFs).
 However, the annihilation effects in the collinear
 approximation exhibit endpoint singularities (ES) for
 charmless mesonic $B$ decays, displaying
 inconsistency of the QCDF formula \cite{qcdf}.

 To deal with ES in convolution integrals,
 many attempts will have been made.

 (1) A phenomenological parameterization of ES in
 annihilation contributions is originally proposed
 by QCDF itself \cite{qcdf}, which ont only
 introduces uncertainties in the QCDF's
 prediction of observables, especially for annihilation
 dominated processes \cite{du}, but also provide
 no constraint on magnitudes of strong phases relevant
 to $CP$ violation.

 (2) ES is removed by separating the physics at different
 momentum scales using the zero-bin subtraction to avoid
 double counting of soft degrees of freedom in SCET
 \cite{prd.76.074002}, while the imaginary part of the
 amplitude is also dropped at the leading power in
 ${\alpha}_{s}{\Lambda}_{\rm QCD}/m_{b}$.

 (3) The infrared finite gluon propagator and running
 coupling constant \cite{prd.26.1453}, or/and Cutkosky
 rules \cite{jmp.1.429} for the quark propagators, are
 used to serve as a natural cutoff , which has already
 been applied to $B$ decays into two mesons
 \cite{yang,su,ijmpa.24.4133}.
 However, it is claimed \cite{ijmpa.24.4133} that different
 predictions on branching ratios can be obtained with
 different solutions of the Schwinger-Dyson equations
 for gluon propagator and coupling constant due to
 different truncations and approximations.

 (4) By keeping the parton transverse momentum $k_{T}$,
 and employing the Sudakov factors to smear the double
 logarithm in QCD radiative corrections and to suppress
 the endpoint contribution of hadron wave functions in
 small transverse momentum region, ES in collinear
 approximation is eliminated with the pQCD approach,
 and the strong phases are perturbatively calculated
 \cite{pqcd}.
 An example is the recently renewed study on the pure
 annihilation decays $B_{s}$ ${\to}$ ${\pi}^{+}{\pi}^{-}$
 and $B_{d}$ ${\to}$ $K^{+}K^{-}$ with the pQCD approach
 \cite{prd.85.094003} which are in good agreement with
 the CDF and LHCb measurements.

 Despite disputes as to which one of above treatments is
 more effective than others, we will study the $B_{s}$ ${\to}$
 ${\eta}_{c}(J/{\psi})D$ decays with the pQCD approach to
 give a rough estimate of their branching ratios.
 Based on $k_{T}$ factorization, the typical expression
 for the decay amplitudes with the pQCD approach can be
 expressed as
  \begin{equation}
 {\int}{\bf d}k\,C(t)H(k,t){\Phi}(k)e^{-S}
  \label{am01}
  \end{equation}
 where $C$, $H$, ${\Phi}$, and $e^{-S}$ are Wilson
 coefficient, hard-scattering kernel, hadron WFs,
 and Sudakov factor, respectively.
 The typical scale $t$ depends on topology and process.
 For convenience, the kinematics variables are described
 by light cone coordinate.
 The momenta of the valence quarks and hadrons in the
 rest frame of the $B_{s}$ meson are defined as follows:
 \begin{eqnarray}
  p_{1}&=&\frac{m_{1}}{\sqrt{2}}(1,1,{\vec{0}}_{\perp}),     \\
  p_{2}&=&({\eta}_{2}^{+},{\eta}_{2}^{-},{\vec{0}}_{\perp}), \\
  p_{3}&=&({\eta}_{3}^{-},{\eta}_{3}^{+},{\vec{0}}_{\perp}), \\
  k_{i}&=&x_{i}p_{i}+(0,0,\vec{k}_{i{\perp}}),               \\
 {\eta_{i}^{\pm}}&=&\frac{E_{i}{\pm}p}{\sqrt{2}}, \\
 {\epsilon}_{2}&=&\frac{1}{m_{2}}({\eta}_{2}^{+},-{\eta}_{2}^{-},{\vec{0}}_{\perp}),
 \end{eqnarray}
 where the subscript $i$ $=$ $1$, $2$, $3$ refers to
 $B_{s}$, ${\eta}_{c}(J/{\psi})$, $D$ meson.
 $k_{i}$, $\vec{k}_{i{\perp}}$, $x_{i}$ are the momentum,
 transverse momentum and longitudinal momentum fraction
 of light quark of meson, respectively.
 ${\epsilon}_{2}$ is the longitudinal polarization vector
 of $J/{\psi}$ meson. In the rest frame of $B_{s}$ meson,
 $E_{i}$ is the energy of particle $i$, and $p$ is the
 common momentum of final state.

 The basic input element in Eq.(\ref{am01})
 --- WFs --- is defined by the nonlocal bilinear
 quark operator matrix element \cite{cc}.
  %%%%% wave function of Bs meson
  \begin{equation}
 {\langle}0{\vert}\bar{b}_{\alpha}(0)s_{\beta}(z)
 {\vert}B_{s}(p_{1}){\rangle}
 =\frac{-i}{\sqrt{2N_{c}}}{\int}{\bf d}^{4}k_{1}
  \,e^{-ik_{1}{\cdot}z} \Big\{ \Big[
  \frac{\not{\!n}_{-}}{\sqrt{2}}{\phi}_{B}^{+}(k_{1})\!
 +\frac{\not{\!n}_{+}}{\sqrt{2}}{\phi}_{B}^{-}(k_{1})
  \Big] \Big(\!\!\not{\!p}_{1}\!+\!m_{B}\!\Big)
 {\gamma}_{5}\Big\}_{{\beta}{\alpha}},
  \label{wf-bs-01}
  \end{equation}
  %%%%% wave function of J/psi meson
  \begin{equation}
 {\langle}J/{\psi}(p_{2}){\vert}\bar{c}_{\alpha}(0)c_{\beta}(z)
 {\vert}0{\rangle}
 =\frac{1}{\sqrt{2N_{c}}}{\int}{\bf d}^{4}k_{2}
  \,{\rm e}^{+ik_{2}{\cdot}z}\!\!\not{\!\epsilon}_{2}
  \Big[m_{J/{\psi}}{\phi}_{\psi}^{L}(k_{2})
 +\!\!\not{\!p}_{2}{\phi}_{\psi}^{t}(k_{2})
  \Big]_{{\beta}{\alpha}},
  \label{wf-jpsi-01}
  \end{equation}
  %%%%% wave function of etac meson
  \begin{equation}
 {\langle}{\eta}_{c}(p_{2}){\vert}\bar{c}_{\alpha}(0)c_{\beta}(z)
 {\vert}0{\rangle}
 =\frac{-i}{\sqrt{2N_{c}}}{\int}{\bf d}^{4}k_{2}
  \,{\rm e}^{+ik_{2}{\cdot}z}\Big\{\!{\gamma}_{5}
  \Big[\!\!\not{p}_{2}{\phi}_{{\eta}_{c}}^{v}(k_{2})
  \!+\!m_{{\eta}_{c}}{\phi}_{{\eta}_{c}}^{s}(k_{2})
  \Big]\Big\}_{{\beta}{\alpha}},
  \label{wf-etac-01}
  \end{equation}
  %%%%% wave function of D meson
  \begin{equation}
 {\langle}D(p_{3}){\vert}\bar{c}_{\alpha}(0)u_{\beta}(z)
 {\vert}0{\rangle}
 =\frac{-i}{\sqrt{2N_{c}}}{\int}{\bf d}^{4}k_{3}
  \,{\rm e}^{+ik_{3}{\cdot}z}\Big[ {\gamma}_{5}
  \Big(\!\!\not{\!p}_{3}\!+\!m_{D}\!\Big){\phi}_{D}(k_{3})
  \Big]_{{\beta}{\alpha}},
  \label{wf-D-01}
  \end{equation}
  where $N_{c}$ $=$ $3$ is the color number.
  $n_{-}$ and $n_{+}$ are null vectors,
  and $n_{+}{\cdot}n_{-}$ $=$ $1$.

  Here, the distribution amplitude of $D$ meson given
  in \cite{prd.67.054028} is used,
  \begin{equation}
 {\phi}_{D}(x)=\frac{f_{D}}{2\sqrt{2N_{c}}}6x\bar{x}
  \Big[1+C_{D}(1-2x)\Big],
  \label{wf-D-02}
  \end{equation}
  where $\bar{x}$ $=$ $1$ $-$ $x$.
  $f_{D}$ is the decay constant. $C_{D}$ is a shape parameter.

  For WFs of ${\eta}_{c}$ and $J/{\psi}$ mesons,
  ${\phi}_{{\eta}_{c}}^{v}$ and ${\phi}_{\psi}^{L}$ are twist-2;
  ${\phi}_{{\eta}_{c}}^{s}$ and ${\phi}_{\psi}^{t}$ are twist-3.
  They can be extracted from the Schr\"{o}dinger state with
  dynamical potentials \cite{cc,epjc.60.107}.
  We will consider two kinds of WFs corresponding to
  harmonic-oscillator and Coulomb potentials.
  Their expressions are listed in \cite{epjc.60.107}.
  One is the harmonic-oscillator (O) type
   \begin{eqnarray}
  {\phi}_{\psi}^{L}(x,b)
  &=&
   \frac{f_{J/{\psi}}}{2\sqrt{2N_{c}}}
   N^{L}_{\psi}x\bar{x}
  {\exp}\Big\{-\frac{m_{c}}{\omega}x\bar{x}
   \Big[\Big(\frac{x-\bar{x}}{2x\bar{x}}\Big)^{2}
 +{\omega}^{2}b^{2} \Big] \Big\}
   \label{wf-jpsi-02},
   \\
  {\phi}_{\psi}^{t}(x,b)
  &=&
   \frac{f_{J/{\psi}}}{2\sqrt{2N_{c}}}
   N^{t}_{\psi}(x-\bar{x})^{2}
  {\exp}\Big\{-\frac{m_{c}}{\omega}x\bar{x}
   \Big[\Big(\frac{x-\bar{x}}{2x\bar{x}}\Big)^{2}
 +{\omega}^{2}b^{2} \Big] \Big\}
   \label{wf-jpsi-03},
   \\
  {\phi}_{{\eta}_{c}}^{v}(x,b)
  &=&
   \frac{f_{{\eta}_{c}}}{2\sqrt{2N_{c}}}
   N^{v}_{{\eta}_{c}}x\bar{x}
  {\exp}\Big\{-\frac{m_{c}}{\omega}x\bar{x}
   \Big[\Big(\frac{x-\bar{x}}{2x\bar{x}}\Big)^{2}
 +{\omega}^{2}b^{2} \Big] \Big\}
   \label{wf-etac-02},
   \\
  {\phi}_{{\eta}_{c}}^{s}(x,b)
  &=&
   \frac{f_{{\eta}_{c}}}{2\sqrt{2N_{c}}}N^{s}_{{\eta}_{c}}
  {\exp}\Big\{-\frac{m_{c}}{\omega}x\bar{x}
   \Big[\Big(\frac{x-\bar{x}}{2x\bar{x}}\Big)^{2}
 +{\omega}^{2}b^{2} \Big] \Big\}
   \label{wf-etac-03}.
   \end{eqnarray}
 The other is the Coulomb (C) type
  \begin{eqnarray}
 {\phi}_{\psi}^{L}(x,b)
 &=&
  \frac{f_{J/{\psi}}}{2\sqrt{2N_{c}}} N^{L}_{\psi}
  \frac{(x\bar{x})^{2}m_{c}b}{\sqrt{1-4x\bar{x}(1-v^{2})}}
  K_{1}(m_{c}b\sqrt{1-4x\bar{x}(1-v^{2})})
  \label{wf-jpsi-05},
  \\
 {\phi}_{\psi}^{t}(x,b)
 &=&
  \frac{f_{J/{\psi}}}{2\sqrt{2N_{c}}} N^{t}_{\psi}
  \frac{(x-\bar{x})^{2}x\bar{x}m_{c}b}{\sqrt{1-4x\bar{x}(1-v^{2})}}
  K_{1}(m_{c}b\sqrt{1-4x\bar{x}(1-v^{2})})
  \label{wf-jpsi-06},
  \\
 {\phi}_{{\eta}_{c}}^{v}(x,b)
 &=&
  \frac{f_{{\eta}_{c}}}{2\sqrt{2N_{c}}} N^{v}_{{\eta}_{c}}
  \frac{(x\bar{x})^{2}m_{c}b}{\sqrt{1-4x\bar{x}(1-v^{2})}}
  K_{1}(m_{c}b\sqrt{1-4x\bar{x}(1-v^{2})})
  \label{wf-etac-05},
  \\
 {\phi}_{{\eta}_{c}}^{s}(x,b)
 &=&
  \frac{f_{{\eta}_{c}}}{2\sqrt{2N_{c}}} N^{s}_{{\eta}_{c}}
  \frac{x\bar{x}m_{c}b}{\sqrt{1-4x\bar{x}(1-v^{2})}}
  K_{1}(m_{c}b\sqrt{1-4x\bar{x}(1-v^{2})})
  \label{wf-etac-06}.
  \end{eqnarray}
 where $f_{J/{\psi}}$ and $f_{{\eta}_{c}}$ are decay constants.
 $m_{c}$ is the mass of $c$ quark.
 $b$ is the conjugate variable of the transverse momentum.
 ${\omega}$ ${\approx}$ $0.6$ GeV and $v$ ${\approx}$ $0.3$
 \cite{epjc.60.107} are shape parameters.
 $N^{L,t}_{\psi}$ and $N^{v,s}_{{\eta}_{c}}$ are the
 normalization constants.
 The normalization conditions are
  \begin{eqnarray}
 {\int}_{0}^{1}{\bf d}x~{\phi}_{\psi}^{L,t}(x,0)
 &=&
  \frac{f_{J/{\psi}}}{2\sqrt{2N_{c}}}
  \label{wf-jpsi-04},
  \\
 {\int}_{0}^{1}{\bf d}x~{\phi}_{{\eta}_{c}}^{v,s}(x,0)
 &=&
  \frac{f_{{\eta}_{c}}}{2\sqrt{2N_{c}}}
  \label{wf-etac-04}.
 \end{eqnarray}

  For WFs of $B_{s}$ meson, there are two scalar
  compositions ${\phi}^{+}_{B}$ and ${\phi}^{-}_{B}$.
  Neglecting three-particle amplitudes, the equation
  of motion for ${\phi}_{B}^{\pm}$ is
  \cite{npb.592.3,plb.523.111}
  \begin{equation}
 {\phi}_{B}^{+}(x)+x\,{\phi}_{B}^{-{\prime}}(x)=0
  \label{eom}.
  \end{equation}
 The relation of Eq.(\ref{eom}) is sometimes referred to
 as ``Wandzura-Wilczek relation'' \cite{plb.72.195}.
 It is helpful in constraining models for WFs, which leads
 to ${\phi}_{B}^{+}(x)$ vanished at the endpoint and
 ${\phi}_{B}^{-}(x)$ $=$ ${\cal O}(1)$ for $x$ ${\to}$ $0$
 \cite{npb.625.239}.
 Here we will investigate three models of WFs for $B_{s}$
 meson. The first one is the exponential (GN) type suggested
 in \cite{prd.55.272}, i.e.,
  \begin{eqnarray}
 {\phi}_{B_{s}}^{{\rm GN}+}(x,b)
 &=&
  \frac{f_{B_{s}}}{2\sqrt{2N_{c}}}N_{\rm GN}^{+}x\,
 {\exp}\Big[-\frac{x\,m_{B_{s}}}{{\omega}_{\rm GN}}\Big]
  \frac{1}{1+(b\;{\omega}_{\rm GN})^{2}}
  \label{wf-bsp-01},
  \\
 {\phi}_{B_{s}}^{{\rm GN}-}(x,b)
 &=&
  \frac{f_{B_{s}}}{2\sqrt{2N_{c}}}N_{\rm GN}^{-}
 {\exp}\Big[-\frac{x\,m_{B_{s}}}{{\omega}_{\rm GN}}\Big]
  \frac{1}{1+(b\;{\omega}_{\rm GN})^{2}}
  \label{wf-bsm-01}.
  \end{eqnarray}
 The second one is the Gaussian (KLS) type proposed in
 \cite{prd.65.014007,prd.74.014027}, i.e.,
  \begin{eqnarray}
 {\phi}_{B_{s}}^{{\rm KLS}+}(x,b)
 &=&
  \frac{f_{B_{s}}}{2\sqrt{2N_{c}}}
  N_{\rm KLS}^{+}x^{2}\bar{x}^{2}\,
 {\exp}\Big[-\frac{1}{2}
  \Big(\frac{x\,m_{B_{s}}}{{\omega}_{\rm KLS}}\Big)^{2}
 -\frac{1}{2}{\omega}_{\rm KLS}^{2}b^{2}\Big]
  \label{wf-bsp-02},
  \\
 {\phi}_{B_{s}}^{{\rm KLS}-}(x,b)
 &=&
  \frac{f_{B_{s}}}{2\sqrt{2N_{c}}}N_{\rm KLS}^{-}
 {\exp}\Big[-\frac{1}{2}{\omega}_{\rm KLS}^{2}b^{2}\Big]\Big\{
 {\exp}\Big[-\frac{1}{2}
  \Big(\frac{x\,m_{B_{s}}}{{\omega}_{\rm KLS}}\Big)^{2}
  \Big]\Big(m_{B_{s}}^{2}\bar{x}^{2}+2{\omega}_{\rm KLS}^{2}\Big)
  \nonumber \\ & &~~~~~~~
 +\sqrt{2{\pi}}m_{B_{s}}{\omega}_{\rm KLS}{\rm Erf}
  \Big(\frac{x\,m_{B_{s}}}{\sqrt{2}{\omega}_{\rm KLS}}\Big)+C_{\rm KLS}\Big\}
  \label{wf-bsm-02},
  \end{eqnarray}
 where the constant $C_{\rm KLS}$ is chosen so that
 ${\phi}_{B_{s}}^{{\rm KLS}-}(1,b)$ $=$ $0$.
 The third one is the KKQT type derived by QCD equation
 of motion and heavy-quark symmetry constraint
 \cite{epjc.28.515,plb.523.111}, i.e.,
  \begin{eqnarray}
 {\phi}_{B_{s}}^{{\rm KKQT}+}(x,b)&=&
  \frac{f_{B_{s}}}{2\sqrt{2N_{c}}}\frac{2x}{{\omega}_{\rm KKQT}^{2}}
 {\theta}(y)J_{0}\Big(m_{B_{s}}b\sqrt{xy}\Big)
  \label{wf-bsp-03}, \\
 {\phi}_{B_{s}}^{{\rm KKQT}-}(x,b)&=&
  \frac{f_{B_{s}}}{2\sqrt{2N_{c}}}\frac{2y}{{\omega}_{\rm KKQT}^{2}}
 {\theta}(y)J_{0}\Big(m_{B_{s}}b\sqrt{xy}\Big)
  \label{wf-bsm-03},
  \end{eqnarray}
 where $y$ $=$ ${\omega}_{\rm KKQT}$ $-$ $x$.

 In Eq.(\ref{wf-bsp-01}---\ref{wf-bsm-03}),
 $f_{B_{s}}$ is the decay constant.
 ${\omega}_{i}$ is the shape parameters.
 The normalization conditions is
  \begin{equation}
 {\int}_{0}^{1}{\phi}_{B_{s}}^{\pm}(x,0){\bf d}x
 =\frac{f_{B_{s}}}{2\sqrt{2N_{c}}}
  \label{wf-bs-02}.
  \end{equation}

 Within the pQCD framework, the Feynman diagrams for $B_{s}$
 ${\to}$ ${\eta}_{c}D$ decay are shown in FIG.\ref{fig01},
 where (a) and (b) are non-factorizable topologies,
 (c) and (d) are factorizable topologies.
 After a straightforward calculation with the master formula
 of Eq.(\ref{am01}), the decay amplitudes
 can be written as follows
  \begin{equation}
 {\cal A}(B_{s}{\to}{\eta}_{c}D)=
  \frac{G_{F}}{\sqrt{2}}V_{ub}^{\ast}V_{cs}
  \sum\limits_{i=a,b,c,d}{\cal A}_{\rm i}
  \label{am03},
  \end{equation}
 The expressions of ${\cal A}_{\rm i}$ are collected in APPENDIX.
 From the expressions, we can clearly see that both
 ${\phi}_{B_{s}}^{+}$ and ${\phi}_{B_{s}}^{-}$
 contribute to the decay amplitudes.
 The branching ratios in the $B_{s}$ meson rest frame can be
 written as:
 \begin{equation}
  {\cal BR}(B_{s}{\to}{\eta}_{c}D)=
   \frac{{\tau}_{B_{s}}}{8{\pi}}
   \frac{p}{m_{B_{s}}^{2}}
  {\vert}{\cal A}(B_{s}{\to}{\eta}_{c}D){\vert}^{2}
   \label{eq:br-01},
 \end{equation}
 where $p$ is the center-of-mass momentum.

 The input parameters in our numerical calculation are
 collected in TABLE. \ref{tab01}.
 If not specified explicitly, we shall take their central
 values as default input.

 Our study show that (1) contributions of FIG.\ref{fig01}
 (a-c) can provide large strong phases, which is
 consistent with pQCD's statement \cite{pqcd}.
 The interference between factorizable diagrams FIG.\ref{fig01}
 (c) and (d) is destructive, which is, by and large, in
 agreement with previous pQCD's estimate (for example, see
 \cite{annth}).
 The strong phase difference between FIG.\ref{fig01}
 (c) and (d) is independent of model of WFs for $B_{s}$
 meson. The main contribution is from nonfactorizable
 diagram FIG.\ref{fig01} (b).
 (2) The dominant contribution is from
 ${\alpha}_{s}/{\pi}$ ${\leq}$ $0.2$ region,
 implying that despite the small phase
 space, these processes are calculated
 with peturbative theory due to hard
 gluon exchange, where the gluon virtuality scales
 as $k_{g}^{2}$ $>$ $(2m_{c})^{2}$.
 (3) There is very strong interference between
 contributions of WFs ${\phi}_{B}^{+}$ and ${\phi}_{B}^{-}$,
 between contributions of twist-2 and twist-3 WFs for
 ${\eta}_{c}(J/{\psi})$ mesons. Contribution with only
 twist-3 WFs for ${\eta}_{c}(J/{\psi})$ mesons (where
 twist-2 WFs is zero and twist-3 WFs is nonzero) is less
 than 30\%.

 Our numerical results are shown in TABLE. \ref{tab02},
 where the first uncertainty comes from the WF shape parameter
 of $B_{s}$ meson, i.e.,
 ${\omega}_{\rm GN}$ $=$ $0.45{\pm}0.10$ GeV in
 Eq.(\ref{wf-bsp-01}---\ref{wf-bsm-01}),
 ${\omega}_{\rm KLS}$ $=$ $0.45{\pm}0.10$ GeV in
 Eq.(\ref{wf-bsp-02}---\ref{wf-bsm-02}) and
 ${\omega}_{\rm KKQT}$ $=$ $0.25{\pm}0.10$ in
 Eq.(\ref{wf-bsp-03}---\ref{wf-bsm-03});
 the second uncertainty comes from the WF shape parameter
 of $J/{\psi}({\eta}_{c})$ meson, i.e.,
 ${\omega}$ $=$ $0.6{\pm}0.1$ GeV in
 Eq.(\ref{wf-jpsi-02}---\ref{wf-etac-03}) and
 $v$ $=$ $0.3{\pm}0.1$ in
 Eq.(\ref{wf-jpsi-05}---\ref{wf-etac-06});
 the third uncertainty comes from the WF shape parameter
 of $D$ meson, i.e., $C_{D}$ $=$ $0.7{\pm}0.1$ in
 Eq.(\ref{wf-D-02});
 the last uncertainty comes from the choice of hard
 scales $(1{\pm}0.1)t_{i}$ in Eq.(\ref{tab}---\ref{tcd}).
 In addition, decay constants $f_{D}$, $f_{J/{\psi}}$,
 $f_{{\eta}_{c}}$, $f_{B_{s}}$ bring some 10\%
 uncertainty to the branching ratio.

 From the numbers in TABLE. \ref{tab02},
 we can clearly see
 (1) branching ratios are sensitive to the choice of
 shape parameter and model of hadronic WFs for $B_{s}$
 and ${\eta}_{c}(J/{\psi})$ mesons, relative to the
 choice of hard scale.
 It is also found that all branching ratios decrease
 with the increasing shape parameter of hadronic WFs
 for $B_{s}$ and ${\eta}_{c}(J/{\psi})$ mesons.
 (2) Due to CKM factors
 ${\vert}V_{cb}^{\ast}V_{us}{\vert}$ $>$
 ${\vert}V_{ub}^{\ast}V_{cs}{\vert}$, there is
 hierarchic structure
 ${\cal BR}(B_{s}{\to}J/{\psi}\overline{D})$
 $>$ ${\cal BR}(B_{s}{\to}J/{\psi}D)$ and
 ${\cal BR}(B_{s}{\to}{\eta}_{c}\overline{D})$
 $>$ ${\cal BR}(B_{s}{\to}{\eta}_{c}D)$.
 Besides, uncertainty (${\sim}$ 30\%) from
 $V_{ub}^{\ast}V_{cs}$ is much larger than that
 (${\sim}$ 5\%) from $V_{cb}^{\ast}V_{us}$.
 (3) Due to $m_{J/{\psi}}$ $>$ $m_{{\eta}_{c}}$
 and the orbital angular momentum $L_{J/{\psi}\overline{D}(D)}$
 $>$ $L_{{\eta}_{c}\overline{D}(D)}$, the phase space for
 $B_{s}$ ${\to}$ $J/{\psi}\overline{D}(D)$ decay is tighter
 than that for $B_{s}$ ${\to}$ ${\eta}_{c}\overline{D}(D)$ decay.
 With the same input, there are relations
 ${\cal BR}(B_{s}{\to}{\eta}_{c}\overline{D})$
 $>$ ${\cal BR}(B_{s}{\to}J/{\psi}\overline{D})$
 and ${\cal BR}(B_{s}{\to}{\eta}_{c}D)$
 $>$ ${\cal BR}(B_{s}{\to}J/{\psi}D)$.
 (4) Branching ratios for $B_{s}$ ${\to}$
 ${\eta}_{c}\overline{D}$, ${\eta}_{c}D$,
 $J/{\psi}\overline{D}$, $J/{\psi}D$ decays
 are the order of $10^{-7}$, $10^{-8}$, $10^{-8}$,
 $10^{-9}$, respectively.

 The corresponding $U$-spin process
 $B_{d}$ ${\to}$ ${\eta}_{c}(J/{\psi})\overline{D}(D)$
 has been studied \cite{prd.64.071501,prd.65.037504,
 prd.73.094006,epjc.59.683}. In \cite{prd.64.071501},
 it is argued that if the intrinsic charm inside $B$
 meson is not much less than 1\%, branching ratio for
 $\overline{B}^{0}(b\bar{d}c\bar{c})$ ${\to}$
 ${\eta}_{c}(J/{\psi})D$ decay is ${\sim}$ $10^{-4}$,
 which is larger than the present experimental upper
 limit $<$ $1.3{\times}10^{-5}$ \cite{pdg2012}.
 Based on collinear factorization scenario, the $B_{d}$
 ${\to}$ ${\eta}_{c}(J/{\psi})D$ decay is investigated
 in \cite{prd.65.037504} with the approach for exclusive
 processes \cite{prd.22.2157}, where narrow ${\delta}$-function
 like WFs are used. The small overlapping among WFs
 results in branching ratio being about $10^{-7}$
 ${\sim}$ $10^{-8}$ \cite{prd.65.037504}.
 This issue is renewed in \cite{prd.73.094006} with
 pQCD approach in the framework of $k_{T}$ factorization.
 By keeping the parton transverse momentum and taking
 the WFs for $c\bar{c}$ final states given in \cite{cc},
 it is found that branching ratio for $B_{d}$ ${\to}$
 ${\eta}_{c}(J/{\psi})D$ decay is bout $10^{-5}$
 ${\sim}$ $10^{-7}$ \cite{prd.73.094006}.
 Considering the final state interactions, branching ratio
 for $B_{d}$ ${\to}$ $J/{\psi}\overline{D}$ is estimated
 to be $10^{-5}$ ${\sim}$ $10^{-6}$ \cite{epjc.59.683}.
 The results in \cite{prd.65.037504,prd.73.094006} have
 similar hierarchic structure due to kinematics and
 dynamics, i.e., ${\cal BR}(B_{d}{\to}{\eta}_{c}D)$
 $>$ ${\cal BR}(B_{d}{\to}J/{\psi}D)$.
 The method used in our study is the same as \cite{prd.73.094006},
 and similar WFs for ${\eta}_{c}(J/{\psi})$ is employed
 (in our study, the small relativistic corrections to the
 WFs are neglected and two types of WFs are considered).
 A consistent estimation is obtained between ours and
 \cite{prd.73.094006}, using the rate
 $\frac{{\cal BR}(B_{s}{\to}c\bar{c}\overline{D})}{{\cal BR}(B_{d}{\to}c\bar{c}\overline{D})}$
 ${\propto}$ $\frac{{\vert}V_{cb}^{\ast}V_{us}{\vert}^{2}}{{\vert}V_{cb}^{\ast}V_{ud}{\vert}^{2}}$
 ${\propto}$ ${\lambda}^{2}$ ${\sim}$ ${\cal O}(10^{-2})$.

 It is well known that the pure annihilation process
 $B_{s}$ ${\to}$ ${\pi}^{+}{\pi}^{-}$ with branching
 ratio ${\sim}$ ${\cal O}(10^{-7})$ \cite{pipi}
 and pure leptonic rare decay $B_{s}$ ${\to}$
 ${\mu}^{+}{\mu}^{-}$ with branching ratio ${\sim}$
 ${\cal O}(10^{-9})$ \cite{mumu}
 have recently been measured at hadron collider,
 due to the fact that there have accumulated much data
 and that detectors sitting at LHC and Tevatron colliders
 have excellent performance on the final charged particles.
 We believe that $B_{s}$ ${\to}$ ${\eta}_{c}(J/{\psi})D$
 decay could be accessible experimentally in the near future,
 because
 (1) their branching ratio is the same order as (sometimes
 larger than) that for $B_{s}$ ${\to}$ ${\pi}^{+}{\pi}^{-}$,
 ${\mu}^{+}{\mu}^{-}$ decays.
 (2) The final $D$ meson can be tagged by charged kaon and/or
 pion, while tracks of both $K^{\pm}$ and ${\pi}^{\pm}$ are
 be clearly seen by sensitive detectors. Besides,
 signal of ${\eta}_{c}(J/{\psi})$ meson is easily identified
 by its narrow peak in the invariant mass distribution.
 For example, the LHCb has measured many $B_{s}$ decay into
 final states containing a charmonium, such as
 $B_{s}$ ${\to}$ $J/{\psi}K^{+}K^{-}$ \cite{arXiv:1302.1213},
 $J/{\psi}\overline{K}^{{\ast}0}$ \cite{prd.86.071102},
 $J/{\psi}f_{0}(980)$ \cite{prl.109.152002},
 $J/{\psi}K_{s}^{0}$ \cite{plb.713.172} ....
 (3) More and more $B_{s}$ data will be accumulated with
 the running of LHC and advancing SuperKEKB.
 It seems to exist a realistic possibility to study
 rare decays with branching ratio ${\sim}$ ${\cal O}(10^{-9})$.

 In summary, we study the pure weak annihilation process $B_{s}$
 ${\to}$ ${\eta}_{c}(J/{\psi})D$ decay with pQCD approach.
 ES disappear as expected by keeping the parton
 transverse momentum.
 The largest uncertainty in our result is
 mainly from QCD's dynamical property of hadron.
 Branching ratio for $B_{s}$ ${\to}$ ${\eta}_{c}(J/{\psi})D$
 decay depends strongly on model of WFs for $B_{s}$ and
 ${\eta}_{c}(J/{\psi})$ meson. There are some other uncertainties
 considered here, such as the high order corrections, the effects
 of final states interaction, and so on. Our estimate show that
 branching ratios for $B_{s}$ ${\to}$ ${\eta}_{c}\overline{D}$,
 ${\eta}_{c}D$, $J/{\psi}\overline{D}$, $J/{\psi}D$ decays
 are the order of $10^{-7}$, $10^{-8}$, $10^{-8}$, $10^{-9}$,
 respectively. They could be measured in the near future.

 \section*{Acknowledgments}
 This work is supported by {\em National Natural Science Foundation
 of China} under Grant Nos. 11147008, U1232101 and 11275057).
 We thanks the referees for their helpful comments.

   \begin{appendix}

   \section{The amplitudes for $B_{s}$ ${\to}$ $J/{\psi}\overline{D}$ decay}
   \label{app01}
   \begin{equation}
  {\cal A}(B_{s}{\to}J/{\psi}\overline{D})
  =\frac{G_{F}}{\sqrt{2}} V_{cb}^{\ast}V_{us}
   \sum\limits_{i=a,b,c,d}{\cal A}_{\rm i}
   \end{equation}
   \begin{eqnarray}
 i{\cal A}_{\rm a}&=&
   \frac{32{\pi}^{2}C_{F}}{\sqrt{2N}}m_{1}
  {\int}_{0}^{1}{\bf d}x_{1}
  {\int}_{0}^{1}{\bf d}x_{2}
  {\int}_{0}^{1}{\bf d}x_{3}
  {\int}_{0}^{\infty}b_{1}{\bf d}b_{1}
  {\int}_{0}^{\infty}b_{2}{\bf d}b_{2}
  {\int}_{0}^{\infty}{\bf d}b_{3}
   \nonumber \\ &{\times}&
  {\alpha}_{s}(t_{a})C_{1}(t_{a})
   e^{-S_{B}-S_{\psi}-S_{D}}
   H_{ab}({\alpha},{\beta}_{a},b_{1},b_{2})
  {\phi}_{D}(x_{3},b_{3}){\delta}(b_{2}-b_{3})
   \nonumber \\ &{\times}&
   \Big\{~
  {\phi}_{B}^{+}(x_{1},b_{1})
  {\phi}_{\psi}^{L}(x_{2},b_{2}){\eta}_{2}^{+} \Big[
   m_{b}{\eta}_{3}^{-}
  +\sqrt{2}{\eta}(x_{1}-x_{2})
  +\sqrt{2}m_{3}^{2}(x_{1}-x_{3})\Big]
   \nonumber \\ & &
 -{\phi}_{B}^{-}(x_{1},b_{1})
  {\phi}_{\psi}^{L}(x_{2},b_{2}){\eta}_{2}^{-} \Big[
   m_{b}{\eta}_{3}^{+}
  +\sqrt{2}{\eta}(x_{1}-x_{2})
  +\sqrt{2}m_{3}^{2}(x_{1}-x_{3})\Big]
   \nonumber \\ & &
 +{\phi}_{B}^{+}(x_{1},b_{1})
  {\phi}_{\psi}^{t}(x_{2},b_{2})m_{2}m_{3}
   \Big[ m_{b}+\frac{1}{2}m_{1}x_{1}
  -\frac{1}{\sqrt{2}}{\eta}_{2}^{+}x_{2}
  -\frac{1}{\sqrt{2}}{\eta}_{3}^{+}x_{3} \Big]
   \nonumber \\ & &
 -{\phi}_{B}^{-}(x_{1},b_{1})
  {\phi}_{\psi}^{t}(x_{2},b_{2})m_{2}m_{3}
   \Big[ m_{b}+\frac{1}{2}m_{1}x_{1}
  -\frac{1}{\sqrt{2}}{\eta}_{2}^{-}x_{2}
  -\frac{1}{\sqrt{2}}{\eta}_{3}^{-}x_{3}
   \Big] \Big\},
   \label{eq:0a}
   \end{eqnarray}
   \begin{eqnarray}
 i{\cal A}_{\rm b}&=&
   \frac{32{\pi}^{2}C_{F}}{\sqrt{2N}}m_{1}
  {\int}_{0}^{1}{\bf d}x_{1}
  {\int}_{0}^{1}{\bf d}x_{2}
  {\int}_{0}^{1}{\bf d}x_{3}
  {\int}_{0}^{\infty}b_{1}{\bf d}b_{1}
  {\int}_{0}^{\infty}b_{2}{\bf d}b_{2}
  {\int}_{0}^{\infty}{\bf d}b_{3}
   \nonumber \\ &{\times}&
  {\alpha}_{s}(t_{b})C_{1}(t_{b})
   e^{-S_{B}-S_{\psi}-S_{D}}
   H_{ab}({\alpha},{\beta}_{b},b_{1},b_{2})
  {\phi}_{D}(x_{3},b_{3}){\delta}(b_{2}-b_{3})
   \nonumber \\ &{\times}&
   \Big\{ \sqrt{2}m_{1}p\,
  {\phi}_{\psi}^{L}(x_{2},b_{2})
   \Big[{\eta}_{3}^{+}{\phi}_{B}^{+}(x_{1},b_{1})
       +{\eta}_{3}^{-}{\phi}_{B}^{-}(x_{1},b_{1})\Big]
        (x_{1}-\bar{x}_{3})
   \nonumber \\ & &
  +m_{2}m_{3}
  {\phi}_{B}^{+}(x_{1},b_{1})
  {\phi}_{\psi}^{t}(x_{2},b_{2})
   \Big[\frac{1}{2}m_{1}\bar{x}_{1}
  -\frac{1}{\sqrt{2}}{\eta}_{2}^{+}x_{2}
  -\frac{1}{\sqrt{2}}{\eta}_{3}^{+}x_{3} \Big]
   \nonumber \\ & &
  -m_{2}m_{3}
  {\phi}_{B}^{-}(x_{1},b_{1})
  {\phi}_{\psi}^{t}(x_{2},b_{2})
   \Big[\frac{1}{2}m_{1}\bar{x}_{1}
  -\frac{1}{\sqrt{2}}{\eta}_{2}^{-}x_{2}
  -\frac{1}{\sqrt{2}}{\eta}_{3}^{-}x_{3}
   \Big] \Big\},
   \label{eq:0b}
   \end{eqnarray}
   \begin{eqnarray}
 i{\cal A}_{\rm c}&=&
   \frac{-8{\pi}^{2}C_{F}}{N}m_{1}pf_{B_{s}}
  {\int}_{0}^{1}{\bf d}x_{2}
  {\int}_{0}^{1}{\bf d}x_{3}
  {\int}_{0}^{\infty}b_{2}{\bf d}b_{2}
  {\int}_{0}^{\infty}b_{3}{\bf d}b_{3}
   \nonumber \\ &{\times}&
  {\alpha}_{s}(t_{c})
   \Big[C_{1}(t_{c})+NC_{2}(t_{c})\Big]
   e^{-S_{\psi}-S_{D}}
   H_{cd}({\alpha},{\beta}_{c},b_{2},b_{3})
  {\phi}_{D}(x_{3},b_{3})
   \nonumber \\ &{\times}&
   \Big\{ {\phi}_{\psi}^{L}(x_{2},b_{2})
   \Big[m_{1}^{2}-(m_{1}^{2}-m_{3}^{2})x_{2}\Big]
      -2m_{2}m_{3}x_{2}{\phi}_{\psi}^{t}(x_{2},b_{2}) \Big\},
   \label{eq:0c}
   \end{eqnarray}
   \begin{eqnarray}
 i{\cal A}_{\rm d}&=&
   \frac{8{\pi}^{2}C_{F}}{N}m_{1}pf_{B_{s}}
  {\int}_{0}^{1}{\bf d}x_{2}
  {\int}_{0}^{1}{\bf d}x_{3}
  {\int}_{0}^{\infty}b_{2}{\bf d}b_{2}
  {\int}_{0}^{\infty}b_{3}{\bf d}b_{3}
   \nonumber \\ &{\times}&
  {\alpha}_{s}(t_{d})
   \Big[C_{1}(t_{d})+NC_{2}(t_{d})\Big]
   e^{-S_{\psi}-S_{D}}
   H_{cd}({\alpha},{\beta}_{d},b_{3},b_{2})
   \nonumber \\ &{\times}&
  {\phi}_{D}(x_{3},b_{3})
  {\phi}_{\psi}^{L}(x_{2},b_{2})
   \Big\{m_{1}^{2}+m_{3}m_{c}
  -(m_{1}^{2}-m_{2}^{2})x_{3}\Big\},
   \label{eq:0d}
   \end{eqnarray}
   %%%%% %%%%% %%%%%
   \section{The amplitudes for $B_{s}$ ${\to}$ $J/{\psi}D$ decay}
   \label{app02}
   \begin{equation}
  {\cal A}(B_{s}{\to}J/{\psi}D)
  =\frac{G_{F}}{\sqrt{2}} V_{ub}^{\ast}V_{cs}
   \sum\limits_{i=a,b,c,d}{\cal A}_{\rm i}
   \end{equation}
   \begin{eqnarray}
 i{\cal A}_{\rm a}&=&
   \frac{32{\pi}^{2}C_{F}}{\sqrt{2N}}m_{1}
  {\int}_{0}^{1}{\bf d}x_{1}
  {\int}_{0}^{1}{\bf d}x_{2}
  {\int}_{0}^{1}{\bf d}x_{3}
  {\int}_{0}^{\infty}b_{1}{\bf d}b_{1}
  {\int}_{0}^{\infty}b_{2}{\bf d}b_{2}
  {\int}_{0}^{\infty}{\bf d}b_{3}
   \nonumber \\ &{\times}&
  {\alpha}_{s}(t_{a})C_{1}(t_{a})
   e^{-S_{B}-S_{\psi}-S_{D}}
   H_{ab}({\alpha},{\beta}_{a},b_{1},b_{2})
  {\phi}_{D}(x_{3},b_{3}){\delta}(b_{2}-b_{3})
   \nonumber \\ &{\times}&
   \Big\{~
  {\phi}_{B}^{+}(x_{1},b_{1})
  {\phi}_{\psi}^{L}(x_{2},b_{2}){\eta}_{3}^{+} \Big[
   \sqrt{2}m_{1}p(x_{1}-x_{3})-m_{b}{\eta}_{2}^{-}\Big]
   \nonumber \\ & &
 +{\phi}_{B}^{-}(x_{1},b_{1})
  {\phi}_{\psi}^{L}(x_{2},b_{2}){\eta}_{3}^{-} \Big[
   \sqrt{2}m_{1}p(x_{1}-x_{3})+m_{b}{\eta}_{2}^{+}\Big]
   \nonumber \\ & &
 -{\phi}_{B}^{+}(x_{1},b_{1})
  {\phi}_{\psi}^{t}(x_{2},b_{2})m_{2}m_{3}
   \Big[\frac{1}{2}m_{1}x_{1}
  -\frac{1}{\sqrt{2}}{\eta}_{2}^{+}x_{2}
  -\frac{1}{\sqrt{2}}{\eta}_{3}^{+}x_{3} \Big]
   \nonumber \\ & &
 +{\phi}_{B}^{-}(x_{1},b_{1})
  {\phi}_{\psi}^{t}(x_{2},b_{2})m_{2}m_{3}
   \Big[ \frac{1}{2}m_{1}x_{1}
  -\frac{1}{\sqrt{2}}{\eta}_{2}^{-}x_{2}
  -\frac{1}{\sqrt{2}}{\eta}_{3}^{-}x_{3}
   \Big] \Big\},
   \label{eq:1a}
   \end{eqnarray}
   \begin{eqnarray}
 i{\cal A}_{\rm b}&=&
   \frac{32{\pi}^{2}C_{F}}{\sqrt{2N}}m_{1}
  {\int}_{0}^{1}{\bf d}x_{1}
  {\int}_{0}^{1}{\bf d}x_{2}
  {\int}_{0}^{1}{\bf d}x_{3}
  {\int}_{0}^{\infty}b_{1}{\bf d}b_{1}
  {\int}_{0}^{\infty}b_{2}{\bf d}b_{2}
  {\int}_{0}^{\infty}{\bf d}b_{3}
   \nonumber \\ &{\times}&
  {\alpha}_{s}(t_{b})C_{1}(t_{b})
   e^{-S_{B}-S_{\psi}-S_{D}}
   H_{ab}({\alpha},{\beta}_{b},b_{1},b_{2})
  {\phi}_{D}(x_{3},b_{3}){\delta}(b_{2}-b_{3})
   \nonumber \\ &{\times}&
   \Big\{ \sqrt{2}{\phi}_{\psi}^{L}(x_{2},b_{2})
   \Big[{\eta}_{2}^{+}{\phi}_{B}^{+}(x_{1},b_{1})
       -{\eta}_{2}^{-}{\phi}_{B}^{-}(x_{1},b_{1})\Big]
   \Big[{\eta}(x_{1}-\bar{x}_{2})
    +m_{3}^{2}(x_{1}-\bar{x}_{3})\Big]
   \nonumber \\ & &
  -m_{2}m_{3}
  {\phi}_{B}^{+}(x_{1},b_{1})
  {\phi}_{\psi}^{t}(x_{2},b_{2})
   \Big[\frac{1}{2}m_{1}\bar{x}_{1}
  -\frac{1}{\sqrt{2}}{\eta}_{2}^{+}x_{2}
  -\frac{1}{\sqrt{2}}{\eta}_{3}^{+}x_{3} \Big]
   \nonumber \\ & &
  +m_{2}m_{3}
  {\phi}_{B}^{-}(x_{1},b_{1})
  {\phi}_{\psi}^{t}(x_{2},b_{2})
   \Big[\frac{1}{2}m_{1}\bar{x}_{1}
  -\frac{1}{\sqrt{2}}{\eta}_{2}^{-}x_{2}
  -\frac{1}{\sqrt{2}}{\eta}_{3}^{-}x_{3}
   \Big] \Big\},
   \label{eq:1b}
   \end{eqnarray}
   \begin{eqnarray}
  {\cal A}_{\rm c}&=&
 -{\cal A}_{\rm d}(B_{s}{\to}J/{\psi}\overline{D}),
   \label{eq:1c} \\
  {\cal A}_{\rm d}&=&
 -{\cal A}_{\rm c}(B_{s}{\to}J/{\psi}\overline{D})
   \label{eq:1d}
   \end{eqnarray}
   %%%%% %%%%% %%%%%
   \section{The amplitudes for $B_{s}$ ${\to}$ ${\eta}_{c}\overline{D}$ decay}
   \label{app03}
   \begin{equation}
  {\cal A}(B_{s}{\to}{\eta}_{c}\overline{D})
  =\frac{G_{F}}{\sqrt{2}} V_{cb}^{\ast}V_{us}
   \sum\limits_{i=a,b,c,d}{\cal A}_{\rm i}
   \end{equation}
   \begin{eqnarray}
  {\cal A}_{\rm a}&=&
   \frac{32{\pi}^{2}C_{F}}{\sqrt{2N}}m_{1}
  {\int}_{0}^{1}{\bf d}x_{1}
  {\int}_{0}^{1}{\bf d}x_{2}
  {\int}_{0}^{1}{\bf d}x_{3}
  {\int}_{0}^{\infty}b_{1}{\bf d}b_{1}
  {\int}_{0}^{\infty}b_{2}{\bf d}b_{2}
  {\int}_{0}^{\infty}{\bf d}b_{3}
   \nonumber \\ &{\times}&
  {\alpha}_{s}(t_{a})C_{1}(t_{a})
   e^{-S_{B}-S_{{\eta}_{c}}-S_{D}}
   H_{ab}({\alpha},{\beta}_{a},b_{1},b_{2})
  {\phi}_{D}(x_{3},b_{3}){\delta}(b_{2}-b_{3})
   \nonumber \\ &{\times}&
   \Big\{~
  {\phi}_{{\eta}_{c}}^{v}(x_{2},b_{2})
  {\phi}_{B}^{+}(x_{1},b_{1}){\eta}_{2}^{+} \Big[
   m_{b}{\eta}_{3}^{-}+\sqrt{2}{\eta}(x_{1}-x_{2})
  +\sqrt{2}m_{3}^{2}(x_{1}-x_{3})\Big]
   \nonumber \\ & &
 +{\phi}_{{\eta}_{c}}^{v}(x_{2},b_{2})
  {\phi}_{B}^{-}(x_{1},b_{1}){\eta}_{2}^{-} \Big[
   m_{b}{\eta}_{3}^{+}+\sqrt{2}{\eta}(x_{1}-x_{2})
  +\sqrt{2}m_{3}^{2}(x_{1}-x_{3})\Big]
   \nonumber \\ & &
 +{\phi}_{{\eta}_{c}}^{s}(x_{2},b_{2})
  {\phi}_{B}^{+}(x_{1},b_{1})m_{2}m_{3}
   \Big[ m_{b}+\frac{1}{2}m_{1}x_{1}
  -\frac{1}{\sqrt{2}}{\eta}_{2}^{+}x_{2}
  -\frac{1}{\sqrt{2}}{\eta}_{3}^{+}x_{3} \Big]
   \nonumber \\ & &
 +{\phi}_{{\eta}_{c}}^{s}(x_{2},b_{2})
  {\phi}_{B}^{-}(x_{1},b_{1})m_{2}m_{3}
   \Big[ m_{b}+\frac{1}{2}m_{1}x_{1}
  -\frac{1}{\sqrt{2}}{\eta}_{2}^{-}x_{2}
  -\frac{1}{\sqrt{2}}{\eta}_{3}^{-}x_{3} \Big] \Big\},
   \label{eq:2a}
   \end{eqnarray}
   \begin{eqnarray}
  {\cal A}_{\rm b}&=&
   \frac{32{\pi}^{2}C_{F}}{\sqrt{2N}}m_{1}
  {\int}_{0}^{1}{\bf d}x_{1}
  {\int}_{0}^{1}{\bf d}x_{2}
  {\int}_{0}^{1}{\bf d}x_{3}
  {\int}_{0}^{\infty}b_{1}{\bf d}b_{1}
  {\int}_{0}^{\infty}b_{2}{\bf d}b_{2}
  {\int}_{0}^{\infty}{\bf d}b_{3}
   \nonumber \\ &{\times}&
  {\alpha}_{s}(t_{b})C_{1}(t_{b})
   e^{-S_{B}-S_{{\eta}_{c}}-S_{D}}
   H_{ab}({\alpha},{\beta}_{b},b_{1},b_{2})
  {\phi}_{D}(x_{3},b_{3}){\delta}(b_{2}-b_{3})
   \nonumber \\ &{\times}&
   \Big\{ \sqrt{2}{\phi}_{{\eta}_{c}}^{v}(x_{2},b_{2})
  \Big[{\eta}_{3}^{+}{\phi}_{B}^{+}(x_{1},b_{1})
      +{\eta}_{3}^{-}{\phi}_{B}^{-}(x_{1},b_{1})\Big]
  \Big[{\eta}(x_{1}-\bar{x}_{3})+m_{2}^{2}(x_{1}-\bar{x}_{2})\Big]
  \nonumber \\ & &
 -{\phi}_{{\eta}_{c}}^{s}(x_{2},b_{2})
  {\phi}_{B}^{+}(x_{1},b_{1})m_{2}m_{3}
   \Big[\frac{1}{2}m_{1}\bar{x}_{1}
  -\frac{1}{\sqrt{2}}{\eta}_{2}^{+}x_{2}
  -\frac{1}{\sqrt{2}}{\eta}_{3}^{+}x_{3}\Big]
   \nonumber \\ & &
 -{\phi}_{{\eta}_{c}}^{s}(x_{2},b_{2})
  {\phi}_{B}^{-}(x_{1},b_{1})m_{2}m_{3}
   \Big[\frac{1}{2}m_{1}\bar{x}_{1}
  -\frac{1}{\sqrt{2}}{\eta}_{2}^{-}x_{2}
  -\frac{1}{\sqrt{2}}{\eta}_{3}^{-}x_{3}\Big] \Big\},
   \label{eq:2b}
   \end{eqnarray}
   \begin{eqnarray}
  {\cal A}_{\rm c}&=&
   \frac{8{\pi}^{2}C_{F}}{N}m_{1}f_{B_{s}}
  {\int}_{0}^{1}{\bf d}x_{2}
  {\int}_{0}^{1}{\bf d}x_{3}
  {\int}_{0}^{\infty}b_{2}{\bf d}b_{2}
  {\int}_{0}^{\infty}b_{3}{\bf d}b_{3}
   \nonumber \\ &{\times}&
  {\alpha}_{s}(t_{c})
   \Big[C_{1}(t_{c})+NC_{2}(t_{c})\Big]
   e^{-S_{{\eta}_{c}}-S_{D}}
   H_{cd}({\alpha},{\beta}_{c},b_{2},b_{3})
  {\phi}_{D}(x_{3},b_{3})
   \nonumber \\ &{\times}&
   \Big\{ \Big[\sqrt{2}{\eta}_{2}^{+}x_{2}-m_{1}\Big]
  \Big[{\eta}_{2}^{+}{\eta}_{3}^{-}
       {\phi}_{{\eta}_{c}}^{v}(x_{2},b_{2})
      +m_{2}m_{3}{\phi}_{{\eta}_{c}}^{s}(x_{2},b_{2})\Big]
  \nonumber \\ &&+
  \Big[\sqrt{2}{\eta}_{2}^{-}x_{2}-m_{1}\Big]
  \Big[{\eta}_{2}^{-}{\eta}_{3}^{+}
       {\phi}_{{\eta}_{c}}^{v}(x_{2},b_{2})
      +m_{2}m_{3}{\phi}_{{\eta}_{c}}^{s}(x_{2},b_{2})\Big] \Big\},
   \label{eq:2c}
   \end{eqnarray}
   \begin{eqnarray}
  {\cal A}_{\rm d}&=&
   \frac{8{\pi}^{2}C_{F}}{N}m_{1}f_{B_{s}}
  {\int}_{0}^{1}{\bf d}x_{2}
  {\int}_{0}^{1}{\bf d}x_{3}
  {\int}_{0}^{\infty}b_{2}{\bf d}b_{2}
  {\int}_{0}^{\infty}b_{3}{\bf d}b_{3}
   \nonumber \\ &{\times}&
  {\alpha}_{s}(t_{d})
   \Big[C_{1}(t_{d})+NC_{2}(t_{d})\Big]
   e^{-S_{{\eta}_{c}}-S_{D}}
   H_{cd}({\alpha},{\beta}_{d},b_{3},b_{2})
  {\phi}_{D}(x_{3},b_{3})
   \nonumber \\ &{\times}&
   \Big\{
   \Big[m_{1}-\sqrt{2}{\eta}_{3}^{+}x_{3}\Big]
   \Big[{\eta}_{3}^{+}{\eta}_{2}^{-}
        {\phi}_{{\eta}_{c}}^{v}(x_{2},b_{2})
      +m_{2}m_{3}{\phi}_{{\eta}_{c}}^{s}(x_{2},b_{2})\Big]
   \nonumber \\ &&+
   \Big[m_{1}-\sqrt{2}{\eta}_{3}^{-}x_{3}\Big]
   \Big[{\eta}_{3}^{-}{\eta}_{2}^{+}
        {\phi}_{{\eta}_{c}}^{v}(x_{2},b_{2})
      +m_{2}m_{3}{\phi}_{{\eta}_{c}}^{s}(x_{2},b_{2})\Big]
   \nonumber \\ &&
    -m_{3}m_{c}E_{2}{\phi}_{{\eta}_{c}}^{v}(x_{2},b_{2})
    -2m_{2}m_{c}E_{3}{\phi}_{{\eta}_{c}}^{s}(x_{2},b_{2})
   \Big\},
   \label{eq:2d}
   \end{eqnarray}
   %%%% %%%%% %%%%%
   \section{The amplitudes for $B_{s}$ ${\to}$ ${\eta}_{c}D$ decay}
   \label{app04}
   \begin{equation}
  {\cal A}(B_{s}{\to}{\eta}_{c}D)
  =\frac{G_{F}}{\sqrt{2}} V_{ub}^{\ast}V_{cs}
   \sum\limits_{i=a,b,c,d}{\cal A}_{\rm i}
   \end{equation}
   \begin{eqnarray}
  {\cal A}_{\rm a}&=&
   \frac{32{\pi}^{2}C_{F}}{\sqrt{2N}}m_{1}
  {\int}_{0}^{1}{\bf d}x_{1}
  {\int}_{0}^{1}{\bf d}x_{2}
  {\int}_{0}^{1}{\bf d}x_{3}
  {\int}_{0}^{\infty}b_{1}{\bf d}b_{1}
  {\int}_{0}^{\infty}b_{2}{\bf d}b_{2}
  {\int}_{0}^{\infty}{\bf d}b_{3}
   \nonumber \\ &{\times}&
  {\alpha}_{s}(t_{a})C_{1}(t_{a})
   e^{-S_{B}-S_{{\eta}_{c}}-S_{D}}
   H_{ab}({\alpha},{\beta}_{a},b_{1},b_{2})
  {\phi}_{D}(x_{3},b_{3}){\delta}(b_{2}-b_{3})
   \nonumber \\ &{\times}&
   \Big\{~
  {\phi}_{{\eta}_{c}}^{v}(x_{2},b_{2})
  {\phi}_{B}^{+}(x_{1},b_{1}){\eta}_{3}^{+} \Big[
   m_{b}{\eta}_{2}^{-}+\sqrt{2}{\eta}(x_{1}-x_{3})
  +\sqrt{2}m_{2}^{2}(x_{1}-x_{2})\Big]
   \nonumber \\ & &
 +{\phi}_{{\eta}_{c}}^{v}(x_{2},b_{2})
  {\phi}_{B}^{-}(x_{1},b_{1}){\eta}_{3}^{-} \Big[
  m_{b}{\eta}_{2}^{+}+\sqrt{2}{\eta}(x_{1}-x_{3})
  +\sqrt{2}m_{2}^{2}(x_{1}-x_{2})\Big]
   \nonumber \\ & &
 +{\phi}_{{\eta}_{c}}^{s}(x_{2},b_{2})
  {\phi}_{B}^{+}(x_{1},b_{1})m_{2}m_{3}
   \Big[ m_{b}+\frac{1}{2}m_{1}x_{1}
  -\frac{1}{\sqrt{2}}{\eta}_{2}^{+}x_{2}
  -\frac{1}{\sqrt{2}}{\eta}_{3}^{+}x_{3} \Big]
   \nonumber \\ & &
 +{\phi}_{{\eta}_{c}}^{s}(x_{2},b_{2})
  {\phi}_{B}^{-}(x_{1},b_{1})m_{2}m_{3}
   \Big[ m_{b}+\frac{1}{2}m_{1}x_{1}
  -\frac{1}{\sqrt{2}}{\eta}_{2}^{-}x_{2}
  -\frac{1}{\sqrt{2}}{\eta}_{3}^{-}x_{3} \Big] \Big\},
   \label{eq:3a}
   \end{eqnarray}
   \begin{eqnarray}
  {\cal A}_{\rm b}&=&
   \frac{32{\pi}^{2}C_{F}}{\sqrt{2N}}m_{1}
  {\int}_{0}^{1}{\bf d}x_{1}
  {\int}_{0}^{1}{\bf d}x_{2}
  {\int}_{0}^{1}{\bf d}x_{3}
  {\int}_{0}^{\infty}b_{1}{\bf d}b_{1}
  {\int}_{0}^{\infty}b_{2}{\bf d}b_{2}
  {\int}_{0}^{\infty}{\bf d}b_{3}
   \nonumber \\ &{\times}&
  {\alpha}_{s}(t_{b})C_{1}(t_{b})
   e^{-S_{B}-S_{{\eta}_{c}}-S_{D}}
   H_{ab}({\alpha},{\beta}_{b},b_{1},b_{2})
  {\phi}_{D}(x_{3},b_{3}){\delta}(b_{2}-b_{3})
   \nonumber \\ &{\times}&
   \Big\{ \sqrt{2}{\phi}_{{\eta}_{c}}^{v}(x_{2},b_{2})
   \Big[{\eta}_{2}^{+}{\phi}_{B}^{+}(x_{1},b_{1})
       +{\eta}_{2}^{-}{\phi}_{B}^{-}(x_{1},b_{1})\Big]
   \Big[{\eta}(x_{1}-\bar{x}_{2})+m_{3}^{2}(x_{1}-\bar{x}_{3})\Big]
   \nonumber \\ & &
 -{\phi}_{{\eta}_{c}}^{s}(x_{2},b_{2})
  {\phi}_{B}^{+}(x_{1},b_{1})m_{2}m_{3}
   \Big[\frac{1}{2}m_{1}\bar{x}_{1}
  -\frac{1}{\sqrt{2}}{\eta}_{2}^{+}x_{2}
  -\frac{1}{\sqrt{2}}{\eta}_{3}^{+}x_{3}\Big]
   \nonumber \\ & &
 -{\phi}_{{\eta}_{c}}^{s}(x_{2},b_{2})
  {\phi}_{B}^{-}(x_{1},b_{1})m_{2}m_{3}
   \Big[\frac{1}{2}m_{1}\bar{x}_{1}
  -\frac{1}{\sqrt{2}}{\eta}_{2}^{-}x_{2}
  -\frac{1}{\sqrt{2}}{\eta}_{3}^{-}x_{3}\Big] \Big\},
   \label{eq:3b}
   \end{eqnarray}
   \begin{eqnarray}
  {\cal A}_{\rm c}&=&-{\cal A}_{\rm d}(B_{s}{\to}{\eta}_{c}\overline{D})
   \label{eq:3c},\\
  {\cal A}_{\rm d}&=&-{\cal A}_{\rm c}(B_{s}{\to}{\eta}_{c}\overline{D})
   \label{eq:3d}
   \end{eqnarray}
   %%%%% %%%%% %%%%%
   \section{Some formula in the decay amplitudes}
   \label{app05}
   The $S_{B}$ ($S_{{\eta}_{c},J/{\psi}}$ and $S_{D}$) in the factor
   $e^{-S_{B}}$ ($e^{-S_{{\eta}_{c},J/{\psi}}}$ and $e^{-S_{D}}$) is
   defined as
   \begin{eqnarray}
   S_{B}(t)&=&s(x_{1}p_{1}^{+},b_{1})
            +2{\int}_{1/b_{1}}^{t}\frac{d{\mu}}{d{\mu}}
              {\gamma}_{q}({\mu}),
   \label{eq:sb} \\
   S_{{\eta}_{c},J/{\psi}}(t)&=&s(x_{2}p_{2}^{+},b_{2})
            +2{\int}_{1/b_{2}}^{t}\frac{d{\mu}}{d{\mu}}
              {\gamma}_{q}({\mu}),
   \label{eq:sjpsi} \\
   S_{D}(t)&=&s(x_{3}p_{3}^{-},b_{3})
            +2{\int}_{1/b_{3}}^{t}\frac{d{\mu}}{d{\mu}}
              {\gamma}_{q}({\mu}),
   \label{eq:sd}
   \end{eqnarray}
  where the quark anomalous dimension ${\gamma}_{q}$
  $=$ $-{\alpha}_{s}/{\pi}$. The expression of $s(Q,b)$
  is given in \cite{pqcd}.
   \begin{eqnarray}
   H_{ab}({\alpha},{\beta},b_{1},b_{2})
   &=& K_{0}(b_{1}\sqrt{{\beta}})
   \Big\{ {\theta}(b_{1}-b_{2}) H_{0}^{(1)}(b_{1}\sqrt{\alpha})
   J_{0}(b_{2}\sqrt{\alpha})+(b_{1}{\leftrightarrow}b_{2})
   \Big\},
   \\
   H_{cd}({\alpha},{\beta},b_{i},b_{j})
   &=& H_{0}^{(1)}(b_{j}\sqrt{\alpha})
   \Big\{ {\theta}(b_{i}-b_{j})K_{0}(b_{i}\sqrt{{\beta}})
   I_{0}(b_{j}\sqrt{{\beta}})+(b_{i}{\leftrightarrow}b_{j})
   \Big\},
   \end{eqnarray}

\begin{figure}[ht]
 \includegraphics[angle=0,width=0.95 \textwidth,bb=65 680 520 750]{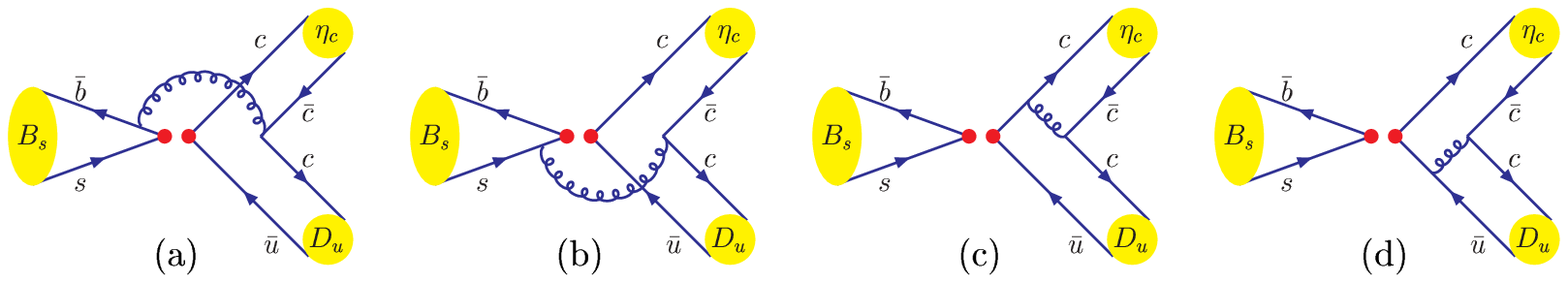}
 \caption{Feynman diagrams for $B_{s}$ ${\to}$ ${\eta}_{c}D$ decay
          within the pQCD framework}
 \label{fig01}
 \end{figure}

 The virtualities of internal gluons (${\alpha}$) and quarks
 (${\beta}_{i}$), and the typical scale $t_{i}$ are defined
 as (where the subscript $i$ $=$ $a$, $b$, $c$, $d$ corresponds
 to the Fig.\ref{fig01})
   \begin{eqnarray}
  {\alpha}&=&\bar{x}_{2}^{2}m_{2}^{2}
            +\bar{x}_{3}^{2}m_{3}^{2}
            +2\bar{x}_{2}\bar{x}_{3}{\eta},
   \label{eq:gg} \\
 -{\beta}_{a}  &=&m_{1}^{2}(x_{1}-x_{2})(x_{1}-x_{3})-m_{b}^{2}
   \nonumber\\ &+&m_{2}^{2}(x_{2}-x_{1})(x_{2}-x_{3})
   \nonumber\\ &+&m_{3}^{2}(x_{3}-x_{1})(x_{3}-x_{2}),
   \label{eq:qa} \\
 -{\beta}_{b}  &=&m_{1}^{2}(x_{1}-\bar{x}_{2})(x_{1}-\bar{x}_{3})
   \nonumber\\ &+&m_{2}^{2}(\bar{x}_{2}-x_{1})(\bar{x}_{2}-\bar{x}_{3})
   \nonumber\\ &+&m_{3}^{2}(\bar{x}_{3}-x_{1})(\bar{x}_{3}-\bar{x}_{2}),
   \label{eq:qb} \\
 -{\beta}_{c}  &=&m_{1}^{2}+x_{2}^{2}m_{2}^{2}
                -x_{2}(m_{1}^{2}+m_{2}^{2}-m_{3}^{2}),
   \label{eq:qc} \\
 -{\beta}_{d}  &=&m_{1}^{2}+x_{3}^{2}m_{3}^{2}-m_{c}^{2}
   \nonumber\\ &-&x_{3}(m_{1}^{2}-m_{2}^{2}+m_{3}^{2}),
   \label{eq:qd}
   \end{eqnarray}
   \begin{eqnarray}
   t_{a(b)}&=&{\max}(\sqrt{\alpha},
               \sqrt{{\vert}{\beta}_{a(b)}{\vert}},
               1/b_{1},1/b_{2},1/b_{3}),
   \label{tab} \\
   t_{c(d)}&=&{\max}(\sqrt{\alpha},
               \sqrt{{\vert}{\beta}_{c(d)}{\vert}},
               1/b_{2},1/b_{3})
   \label{tcd}.
   \end{eqnarray}
   \end{appendix}

  \begin{table}[htb]
  \caption{input parameters for $B_{s}$ ${\to}$ ${\eta}_{c}(J/{\psi})D$ decay}
  \label{tab01}
  \begin{ruledtabular}
  \begin{tabular}{l|cc}
  \multicolumn{1}{c|}{parameter} & value & reference \\ \hline
  mass of $B_{s}$ meson & $m_{B_{s}}$ $=$ $5366.77{\pm}0.24$ MeV         & \cite{pdg2012} \\
  mass of $J/{\psi}$ meson & $m_{J/{\psi}}$ $=$ $3096.916{\pm}0.011$ MeV & \cite{pdg2012} \\
  mass of $\eta_{c}$ meson & $m_{\eta_{c}}$ $=$ $2981.0{\pm}1.1$ MeV     & \cite{pdg2012} \\
  mass of $D$ meson & $m_{D}$ $=$ $1864.86{\pm}0.13$ MeV                 & \cite{pdg2012} \\
  mass of $b$ quark & $m_{b}$ $=$ $4.18{\pm}0.03$ GeV                    & \cite{pdg2012} \\
  mass of $c$ quark & $m_{c}$ $=$ $1.275{\pm}0.025$ GeV                  & \cite{pdg2012} \\
  lifetime of $B_{s}$ meson & ${\tau}_{B_{s}}$ $=$ $1.497{\pm}0.015$ ps  & \cite{pdg2012} \\
  decay constant of $D$ meson & $f_{D}$ $=$ $206.7{\pm}8.9$ MeV          & \cite{pdg2012} \\
  decay constant of $J/{\psi}$ meson & $f_{J/{\psi}}$ $=$ $405{\pm}6$ MeV  & \cite{prd.86.094501} \\
  decay constant of $\eta_{c}$ meson & $f_{{\eta}_{c}}$ $=$ $394.7{\pm}2.4$ MeV & \cite{prd.82.114504} \\
  decay constant of $B_{s}$ meson & $f_{B_{s}}$ $=$ $227.6{\pm}5.0$   MeV  & \cite{fbs} \\
  Wolfenstein parameters & $A$ $=$ $0.811^{+0.022}_{-0.012}$               & \cite{pdg2012} \\
                         & ${\lambda}$ $=$ $0.22535{\pm}0.00065$           & \cite{pdg2012} \\
                         & $\bar{\rho}$ $=$ $0.131^{+0.026}_{-0.013}$      & \cite{pdg2012} \\
                         & $\bar{\eta}$ $=$ $0.345^{+0.013}_{-0.014}$      & \cite{pdg2012}
  \end{tabular}
  \end{ruledtabular}
  \end{table}

 \begin{table}[htb]
 \caption{Branching ratio for $B_{s}$ ${\to}$
  ${\eta}_{c}(J/{\psi})D$ decay with different models
  of WFs.}
 \label{tab02}
 \begin{ruledtabular}
 \begin{tabular}{l|c|ccc}
 \multicolumn{2}{c}{} & GN & KLS & KKQT  \\ \hline
 ${\cal BR}(B_{s}{\to}J/{\psi}\overline{D}){\times}10^{8}$
  & O
  & $1.54^{+0.94+0.61+0.13+0.05}_{-0.61-0.34-0.12-0.05}$
  & $3.02^{+0.84+0.96+0.11+0.02}_{-0.91-0.62-0.08-0.03}$
  & $2.92^{+1.63+0.96+0.08+0.05}_{-1.91-0.61-0.07-0.07}$
  \\
  & C
  & $2.83^{+1.02+3.56+0.09+0.25}_{-0.74-1.73-0.06-0.33}$
  & $4.37^{+0.72+4.76+0.19+0.31}_{-0.89-2.41-0.16-0.42}$
  & $4.00^{+1.58+4.88+0.16+0.30}_{-2.52-2.39-0.14-0.42}$
  \\ \hline
 ${\cal BR}(B_{s}{\to}J/{\psi}D){\times}10^{9}$
  & O
  & $4.34^{+1.37+0.27+0.08+0.31}_{-1.40-0.16-0.07-0.33}$
  & $6.28^{+0.63+0.32+0.17+0.38}_{-0.68-0.22-0.14-0.41}$
  & $7.09^{+0.32+1.00+0.12+0.42}_{-1.84-0.66-0.09-0.46}$
  \\
  & C
  & $ 7.08^{+0.65+1.45+0.07+0.36}_{-1.55-0.98-0.05-0.33}$
  & $ 9.08^{+0.61+2.08+0.27+0.35}_{-1.54-1.30-0.21-0.30}$
  & $15.07^{+0.78+5.72+0.10+0.50}_{-7.15-3.48-0.03-0.45}$
  \\ \hline
 ${\cal BR}(B_{s}{\to}{\eta}_{c}\overline{D}){\times}10^{7}$
  & O
  & $4.09^{+0.87+0.67+0.31+0.07}_{-0.84-0.49-0.27-0.09}$
  & $5.30^{+0.55+0.89+0.36+0.09}_{-0.68-0.64-0.31-0.11}$
  & $5.36^{+0.87+0.89+0.37+0.07}_{-1.58-0.64-0.32-0.09}$
  \\
  & C
  & $5.36^{+1.14+3.57+0.17+0.19}_{-1.05-1.98-0.11-0.24}$
  & $7.22^{+0.63+5.02+0.16+0.24}_{-0.98-2.74-0.07-0.31}$
  & $7.24^{+1.14+4.77+0.19+0.21}_{-2.74-2.65-0.10-0.27}$
  \\ \hline
 ${\cal BR}(B_{s}{\to}{\eta}_{c}D){\times}10^{8}$
  & O
  & $ 9.48^{+1.08+0.21+0.12+0.08}_{-1.43-0.26-0.10-0.00}$
  & $11.43^{+0.06+0.21+0.27+0.07}_{-0.42-0.26-0.24-0.00}$
  & $12.46^{+0.02+0.58+0.22+0.09}_{-1.95-0.55-0.19-0.00}$
  \\
  & C
  & $ 7.96^{+0.29+0.74+0.24+0.09}_{-0.86-0.66-0.22-0.01}$
  & $ 9.39^{+0.34+1.30+0.42+0.05}_{-1.02-0.90-0.38-0.00}$
  & $11.93^{+0.15+2.31+0.33+0.10}_{-3.43-1.65-0.28-0.00}$
 \end{tabular}
 \end{ruledtabular}
 \end{table}

 \end{document}